\documentclass[showpacs,preprintnumbers,amsmath,amssymb,superscriptaddress]{revtex4}
\usepackage{graphicx}
\usepackage{aas_macros}
\usepackage{bm}

\begin{document}
%
\title{A Test for the Zero Mean Hypothesis in Cosmology}
\author{Kiyotomo Ichiki}
\email{ichiki@a.phys.nagoya-u.ac.jp}
\affiliation{%
Kobayashi-Maskawa Institute for the Origin of Particles and
the Universe, Nagoya University, Chikusa-ku, Nagoya, 464-8602, Japan
}
\affiliation{%
Department of Physics and Astrophysics, Nagoya University, Nagoya
464-8602, Japan
}

\date{\today} \preprint{}
\begin{abstract}
 One working hypothesis on which analyses of cosmological data are based
 is the zero ensemble mean hypothesis, which is related to the
 statistical homogeneity of cosmological perturbations. This hypothesis,
 however, should be tested by observational data in the current era of
 precision cosmology. Herein, we test the hypothesis by analyzing
 recent, foreground-reduced cosmic microwave background (CMB) maps,
 combining the spherical harmonic coefficients of the masked CMB
 temperature anisotropies in such a way that the combined variables can
 be treated as statistically independent samples.  We find evidence
 against the zero mean hypothesis in two particular ranges of
 multipoles, with significance levels of $2.5 \sigma$ and $3.1 \sigma$
 in the multipole ranges of $\ell \approx 61$-$86$ and $213$-$256$,
 respectively, for both the Planck and Wilkinson Microwave Anisotropy
 Probe maps. The latter signal is consistent with our previous result
 found by using brute-force Monte-Carlo simulations. However, within the
 method employed in this paper we conclude that the zero mean hypothesis
 is consistent with the current CMB data on the basis of Stouffer's
 weighted Z statistics, which takes multiple testing into account.
\end{abstract}
\maketitle

\section{Introduction}
Recent precise measurements of anisotropies of the cosmic microwave
background (CMB) as well as a number of probes of the large scale structure
(LSS) of the universe have led us to the standard, concordant model of
cosmology. In the standard cosmological model, the universe contains
small density fluctuations on top of otherwise flat, homogeneous, and
isotropic space-time. The density fluctuations are thought to be generated
through quantum fluctuations in the accelerating expansion phase in the
early universe, i.e., inflation. An important diagnostic characteristic of
inflation models is that they predict statistically homogeneous and
isotropic Gaussian fluctuations with a near scale-invariant power
spectrum (for a review, see \cite{2009arXiv0907.5424B}). 

Among these features, the Gaussianity and approximate scale invariance
have been intensively tested by a number of observations and verified
with high significance
\cite{2013arXiv1303.5076P,2013arXiv1303.5084P,2013arXiv1303.5082P,2013ApJS..208...19H},
while the statistical homogeneity and isotropy have been less tested and
often assumed implicitly in cosmological analyses
\cite{2011RSPTA.369.5115M}. Recently, the test of statistical isotropy
has attracted much attention
\cite{2004MNRAS.349..313P,2004ApJ...605...14E,2009ApJ...699..985H,2009ApJ...704.1448H,2008PhRvD..78f3531B},
after the authors of \cite{2004ApJ...605...14E} found hints for the
breaking of statistical isotropy in the Wilkinson Microwave Anisotropy
Probe (WMAP) CMB anisotropy data (see also, \cite{2007ApJ...660L..81E,2013ApJ...773L...3A}). The existence of the statistical
anisotropy has been confirmed by Planck data \cite{2013arXiv1303.5083P},
and more recently Akrami et al. have found the statistical anisotropy
at the $3.3\sigma$ level by
measuring the local variance, using the 1000 available Planck Full Focal
Plane simulations \cite{2014ApJ...784L..42A}.
On the other hand, no evidence has been found in a sample of luminous red galaxies
observed by the Sloan Digital Sky Survey \cite{2010JCAP...05..027P}.

In this paper, we test statistical homogeneity using recent fullsky CMB
temperature maps provided by the WMAP and Planck
satellites. Specifically, we test the null hypothesis that the means of
cosmological perturbations are zero in spherical harmonic space, which
corresponds to the usual Fourier space in three-dimensional space. It is
understood that the zero mean hypothesis is related to statistical
homogeneity as follows \cite{2011JCAP...03..048A}. We usually assume
that because of the cosmological principle, perturbation variables, such as
the CMB temperature, can be decomposed into a space-independent background
value and perturbations, as $T=T_0(t)+\delta T(t,\vec{x})$ with
\begin{equation}
\left<\delta T(t,\vec{x})\right>=0~,
\label{eq:1}
\end{equation}
where the angle brackets denote an ensemble average.
Note that this decomposition can be done only if the expectation value of $T$
is constant, i.e., $\left<T(t,\vec{x})\right>=\mbox{const.}$ with fixed
time.  It is always possible to satisfy Eq.~(\ref{eq:1}) even if the
expectation is not constant, but in that case the background temperature
cannot be space-independent.  The condition that this expectation value is
constant is equivalent to the statistically homogeneous (stationary)
condition of the mean,
\begin{equation} 
\left<\delta T(t,\vec{x})\right>=
\left<\delta T(t,\vec{x}+\vec{X})\right>
\hspace{5mm} {\forall}\vec{X}\in R^3~,
\end{equation}
where the time coordinate $t$ is defined for the background
temperature $T$ to be homogeneous.
Hence the zero mean hypothesis is equivalent to the statistical
homogeneity of the mean. The temperature anisotropy can be written, in
the linearized theory, as
\begin{equation}
 \delta T(\vec{x},\hat{n},t) = \int \frac{d^3 k}{(2\pi)^3}{\cal
  T}(\vec{k},\hat{n})\phi(\vec{k})e^{i\vec{k}\cdot \vec{x}}~,
\label{eq:temp}
\end{equation}
where $\phi(\vec{k})$ are the Fourier modes of the initial density
perturbations and ${\cal T}(\vec{k},\hat{n})$ is the linear transfer
function that relates the initial density perturbations to the currently 
observed temperature anisotropies. Because the temperature
fluctuations are observed on the sphere, it is common practice to
express them with real spherical harmonic coefficients 
\begin{equation}
 a_{\ell m}(\vec{x})=\int d^2\hat{n} \delta T(\vec{x},\hat{n},t)R_{\ell m}(\hat{n})~,
\end{equation}
where  $R_{\ell m}$ is the real set of spherical harmonics.
Because the transfer function in Eq.~(\ref{eq:temp}) is
completely determined by the cosmological perturbation theory given a cosmological model, the zero
mean condition $\left<\delta T(t,\vec{x})\right>=0$ is equivalent to the
conditions $\left<\phi(\vec{k})\right>=\left<a_{\ell m}(\vec{x})\right>=0$. In
the following, we test whether the condition $\left<a_{\ell m}\right>=0$
is satisfied using the latest CMB data, replacing the ensemble
average with the directional average assuming the statistical
homogeneity and isotropy in the mean.

Studies on the zero mean hypothesis using CMB data can be found in
Refs.~\cite{2011JCAP...03..048A} and \cite{2012PhRvD..85f3001K}. One
complication when testing the zero mean hypothesis with CMB data is the
existence of the mask, which suppresses foreground contamination but
generates correlations between the samples. Picon analyzed CMB data by
constructing  $v$-vectors that disentangle the correlations
\cite{2011JCAP...03..048A}, while Kashino et
al. utilized Monte-Calro 
simulations to take into account the correlations in the samples \cite{2012PhRvD..85f3001K}. In
this paper we build on the work of \cite{2011JCAP...03..048A}, and
extend the analysis toward higher multipoles using the latest CMB data
from Planck.

To assess any statistical properties of the foreground-reduced CMB maps
strictly, one needs to use specific simulations.  These simulations
directly reflect uncertainties in the foreground cleaning methods,
residuals of the antenna beam shape, anisotropy of the noise and others
\cite{2004MNRAS.349..313P}. 
For the Planck experiment, we adopt hundred CMB and noise
simulations produced by the Planck collaboration \footnote{http://wiki.cosmos.esa.int/planckpla/index.php/Simulation\_data}, and use these
simulations to address these issues. In addition we compare the WMAP and
Planck maps, which have 
different residuals of the antenna beam shape. We also compare the different
foreground-reduced Planck maps, which have different uncertainties in
the foreground cleaning methods. We will see that these maps give
consistent results.

\section{Method}
Herein we summarize the method developed by \cite{2011JCAP...03..048A}.
Observed temperature fluctuations, $\delta T(\hat{n})_{\rm obs}$, involve
a convolution with the detector beam window $B$ and the pixel smoothing
kernel $K$ and can be expressed as 
\begin{equation}
 \delta T^{\rm obs}(\hat{n})=M(\hat{n})\left(
K\ast\left[B\ast \delta T^{\rm CMB}(\hat{n})+N(\hat{n})\right]
\right)~,
\end{equation}
where $\delta T_{\rm CMB}(\hat{n})$ is the CMB signal we want to
estimate, $N(\hat{n})$ is instrument noise and $M(\hat{n})$ is the
mask. Here we have omitted the foreground assuming that 
$M(\hat{n})$ can successfully mask the foreground contaminated
regions. In spherical harmonic space, the above equation is
expressed as
\begin{equation}
 a^{\rm obs}_{\ell m}=\sum M_{\ell m; \ell^\prime m^\prime}
  a^{\rm full\, sky}_{\ell^\prime m^\prime}~,
\end{equation}
where $a^{\rm full\, sky}_{\ell m}=K_\ell B_\ell a^{\rm CMB}_{\ell
m}+K_\ell n_{\ell m}$ consists of the spherical
harmonic coefficients of the sky, including signal and noise, and $M_{\ell m;
\ell^\prime m^\prime}$ is the mask-coupling matrix. Because instrumental
noise is expected to be well described by a Gaussian distribution as
shown in the WMAP \cite{2011ApJS..192...14J} and Planck papers, ensemble averages of $a^{\rm obs}_{\ell
m}$ are zero if those of $a^{\rm CMB}_{\ell m}$ are zero.
Considering an axisymmetric mask that satisfies $M_{\ell m; \ell^\prime
m^\prime}=M_{\ell m; \ell^\prime m}\delta_{mm^\prime}$ and 
using matrix notation for a fixed $m$, the above equation can be
expressed as 
\begin{equation}
 \vec{a}_m^{\, \rm obs} = {\bm M}\cdot \vec{a}_m^{\, \rm full\, sky}~,
\label{eq:3}
\end{equation}
where the dot represents inner product over multipoles $\ell$.
To remove the effect of the mask from the observed spherical harmonic
coefficients and disentangle the coupling, consider $m$-independent
$v$-vectors that satisfy the relation
\begin{equation}
 \vec{v}^{\,t}=\vec{v}^{\,t} {\bm M}~,
\label{eq:vec{v}}
\end{equation}
where
\begin{equation}
v_{\ell m} = \begin{cases}
    v_\ell & {\rm for}~ |m|\leq \ell_{\rm min} ~\mbox{and}~ \ell_{\rm min} \leq \ell \leq \ell_{\rm max}\\
    0 &  \mbox{(otherwise)} ~.
  \end{cases} 
\end{equation}
Let us construct a variable $d_m$ as the dot product of $\vec{v}$ and
$\vec{a}^{\,\rm obs}_m$ 
\begin{equation}
 d_m \equiv \vec{v}^{\,t}\cdot \vec{a}^{\,\rm obs}
 = \vec{v}^{\,t} {\bm M} \vec{a}_m^{\, \rm full\, sky}
 = \vec{v}^{\,t} \cdot \vec{a}_m^{\, \rm full\, sky}~,
\label{eq:dm}
\end{equation}
for $|m|<m_{\rm max}$.
The new stochastic variable $d_m$ has the following properties:
\begin{itemize}
 \item[1.] Foreground insensitive, because we work on $a^{\rm obs}_{\ell
       m}$, the spherical harmonic coefficients of the masked sky
 \item[2.] Statistically independent, because they are constructed as a
       linear combination of statistically independent variables $a^{\rm
       CMB}_{\ell m}$ and $n_{\ell m}$
 \item[3.] Gaussian with zero mean, if $a^{\rm CMB}_{\ell m}$ and
	   $n_{\ell m}$ are as well
 \item[4.] Having $m$-independent variance, where $\sigma^2 = \sum_{\ell_{\rm
       min}}^{\ell_{\rm max}} K_\ell (B_\ell^2 C_\ell+N_\ell)v_\ell^2$.
\end{itemize}
Owing to these properties, we can formulate a simple statistical test of
the zero mean hypothesis. In our test below, we estimate $\sigma^2$
directly from the data.

To obtain the $v$-vectors, it is convenient to work in pixel
 space. In pixel space, Eq.(\ref{eq:vec{v}}) is written as
 $(1 - M(\hat{n}))v(\hat{n})=0$. Substituting $v(\hat{n})=\sum v_{\ell
 m}Y_{\ell m}(\hat{n})$ and defining the matrix 
\begin{equation}
D_{i\ell}=\left(1-M(\hat{n}_i)\right)
\sum_{|m|\leq \ell_{\rm min}}Y_{\ell m}(\hat{n}_i)~,
\end{equation}
where $i$ runs over all pixels and $\ell_{\rm min}\leq \ell \leq
\ell_{\rm max}$,  we can rewrite the system of equations in Eq.~(\ref{eq:vec{v}}) as
\begin{equation}
 {\bm D}\vec{v}=0~,
\end{equation}
or, in component notation, $\sum_\ell D_{i\ell} v_\ell=0$.
The dimension of the matrix $D$ is $(\ell_{\rm max} - \ell_{\rm
min}+1,N_{\rm pix})$,
where we have used the Healpix pixelization scheme with $N_{\rm
side}=256$, and therefore $N_{\rm pix}=786432$. We find an approximate
solution of this system of equations using singular value
decomposition (SVD). The SVD of the matrix $D$ is expressed as
\begin{equation}
 {\bm D} = {\bm U}{\bm \Sigma}{\bm V}^T~,
\end{equation}
where ${\bm U}{\bm U}^T={\bm V}{\bm V}^T={\bm I}$. The columns of ${\bm
U}$ and ${\bm V}$ are orthogonal eigenvectors of ${\bm D}{\bm D}^T$ and
${\bm D}^T{\bm D}$, respectively, and ${\bm \Sigma}$ is a diagonal
matrix containing the singular values of the matrix ${\bm D}$ in
descending order. We choose
the vector $\vec{v}$ to be the last right singular vector, so that
\begin{equation}
 |{\bm D}\vec{v}|^2 = \Sigma_{\rm last}^2~.
\end{equation}
The last singular value, $\Sigma_{\rm last}$, should be small but
non-zero, and therefore our solution is only approximate. Following
\cite{2011JCAP...03..048A}, we choose the
binning of multipoles so that the last singular value divided by the
norms of the mask and the sky encoded in $\vec{v}$ is sufficiently small
($\lesssim 10^{-8}$). 

\section{Result}
\subsection{test using a simple statistic}
\begin{figure}[h]
\centering
\includegraphics[width=0.9\textwidth]{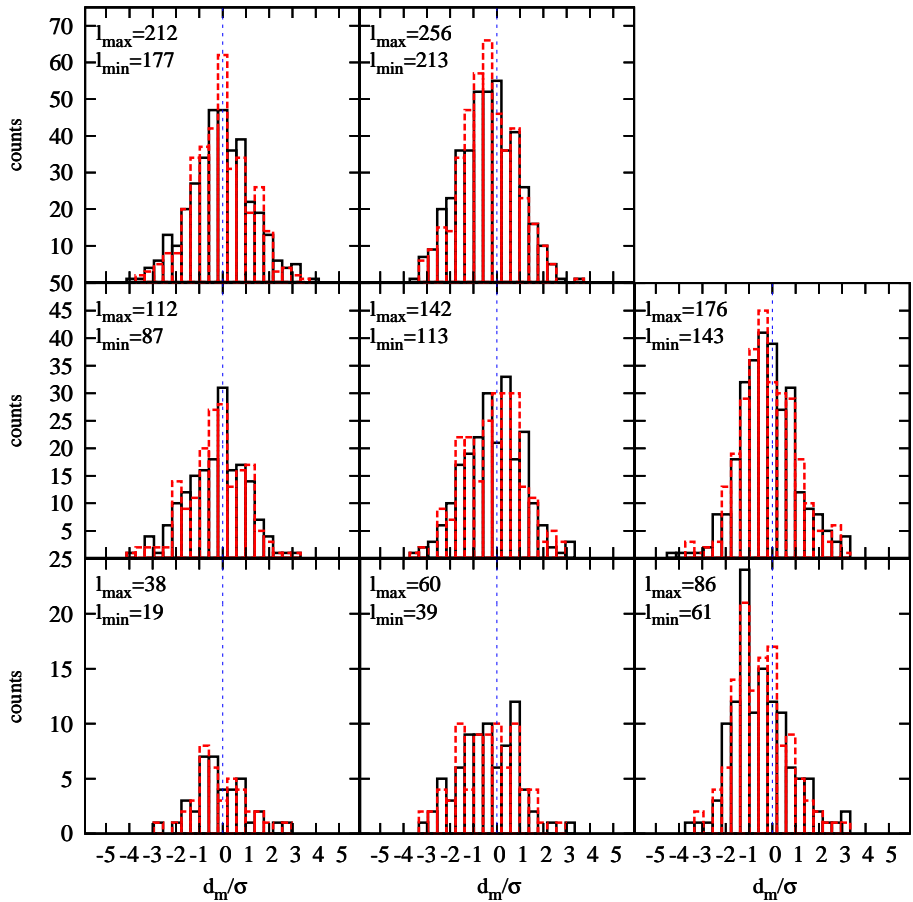}
\caption{Histograms of stochastic variable $d_m$ normalized by the sample
 variance. Binning in $\ell$ space has been done and is shown in the
 figure. If the $a_{\ell m}$ of the CMB follow a Gaussian distribution
 with a zero mean, so does the variable $d_m$. The red histograms were
 obtained from the 
 WMAP map, and the black from Planck map.}
\label{fig:hist}
\end{figure}

We tested the zero mean hypothesis with the stochastic variable $d_m$,
which is constructed as a linear combination of $a_{\ell m}$ given by
Eq.~(\ref{eq:dm}). To perform the inner product in Eq.~(\ref{eq:dm}), we
divide the multipoles into bins \cite{2011JCAP...03..048A}, and the
ranges of these bins are
shown in Fig.~\ref{fig:hist}. In the figure, we show histograms of the
variable $d_m$, constructed from the WMAP (red) and Planck (black) maps.
The variable $d_m$ is normalized by the sample variance $\sigma$.  It
is evident from the figure that in the examined multipole range the
Planck and WMAP maps give consistent results. 
The means of the
distributions are consistent with zero, except for possible deviations for the multipole
ranges of 
$\ell \approx 61$ - $86$ and $\ell \approx 213$ - $256$.

\begin{figure}[h]
\centering
\includegraphics[width=0.45\textwidth]{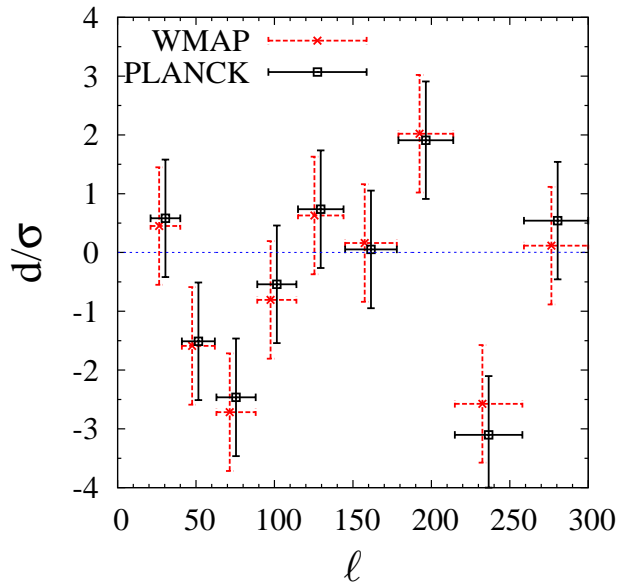}
\caption{Deviations from the zero mean hypothesis from WMAP nine-year
 data (red
 dashed) and Planck SMICA (black solid) maps. The vertical error 
 bars are $\pm \sigma$, while the horizontal error bars are the width of
 the binning. }
\label{fig:z-score}
\end{figure}

In Fig.~\ref{fig:z-score}, we depict the result of the test showing how
many sigmas the observed data deviate from the zero mean.  Here we
perform a simple test assuming Gaussian statistics as follows. We estimate
the mean from the sample by computing $\bar{d}=\sum_m d_m / (2\ell_{\rm
min}+1)$ and then obtain the error in the estimate of the mean from the
formula
\begin{equation}
\sigma (\bar{d})= \frac{\sigma}{\sqrt{2\ell_{\rm min}+1}}~,\label{eq:3.15}
\end{equation}
where $\sigma^2 \equiv \sum (d_m-\bar{d})^2/(2\ell_{\rm min})$ is the
sample variance. The Z-scores in the figure are simply defined by
$Z=\bar{d}/\sigma (\bar{d})$ for each multipole bin.

There are hints of deviations in the ranges around
$\ell\approx 70$ and $\ell \approx 230$. Significance levels are
$2.5 \sigma$ for the former and $3.1\sigma$ for the latter. For the
latter signal, the
significance is slightly larger for the Planck map. The other
multipole ranges are consistent with the zero mean hypothesis.

\subsection{test using Planck simulations}
In the simple test presented above we assume symmetric instrumental
beams and isotropic noise. However, in actual experiments such as the
WMAP and Planck ones, beams are not perfectly symmetric, and noise is
anisotropic due to the scan strategy. In fact, asymmetric instrumental
beams can introduce statistical artifacts (e.g.,
\cite{2009ApJ...707..343W}), and anisotropic scan strategy can break the
zero mean assumption in the noise. To see whether these effects change
the results in the previous section, we adopt the hundred CMB and noise
sky simulations produced by the Planck collaboration and apply the same
method to these sky maps.  In these simulations they have taken into
account the scan strategy, the instrumental performance, and the noise
property of the Planck experiment.

\begin{figure}[h]
\begin{minipage}[m]{0.49\textwidth}
\centering
\includegraphics[width=1.0\textwidth]{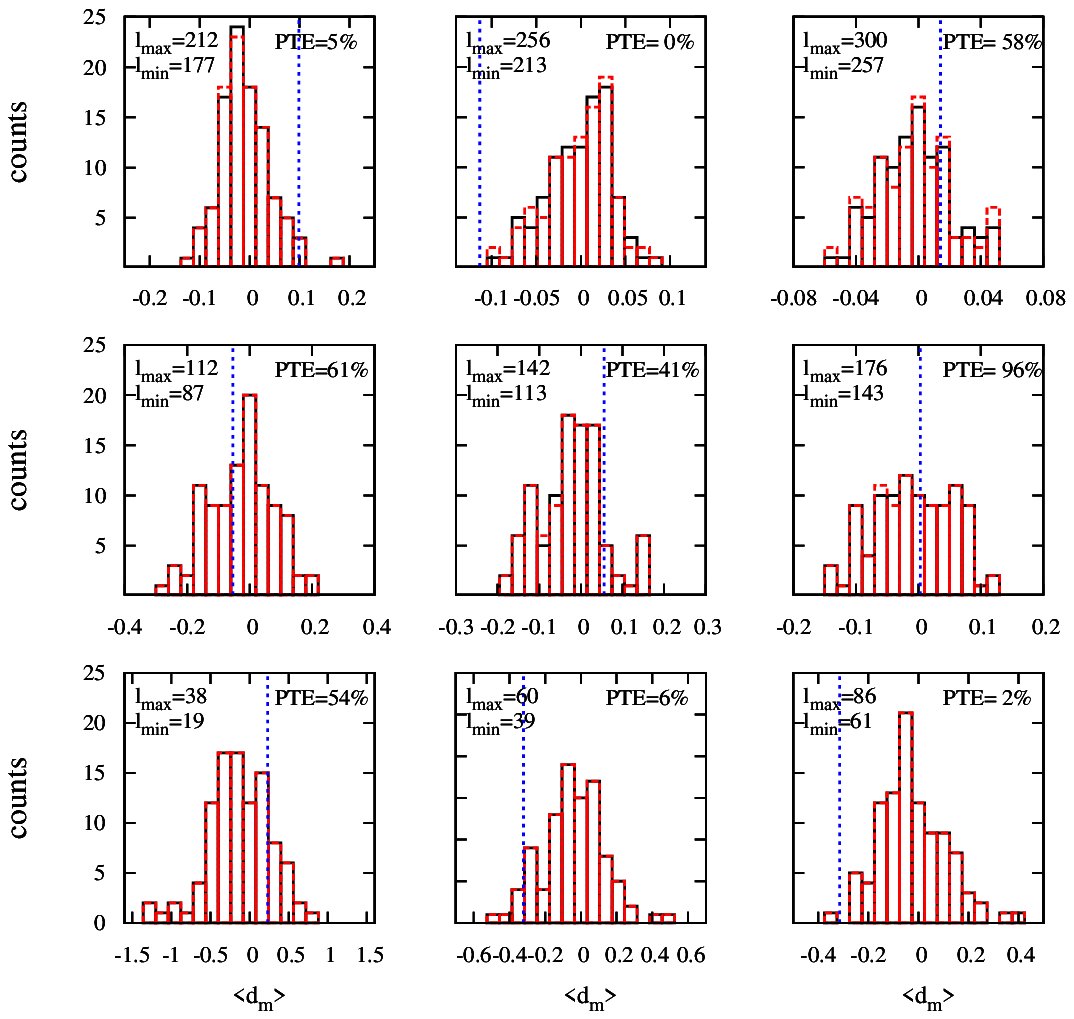}
\end{minipage}
\begin{minipage}[m]{0.49\textwidth}
\centering
\includegraphics[width=0.8\textwidth]{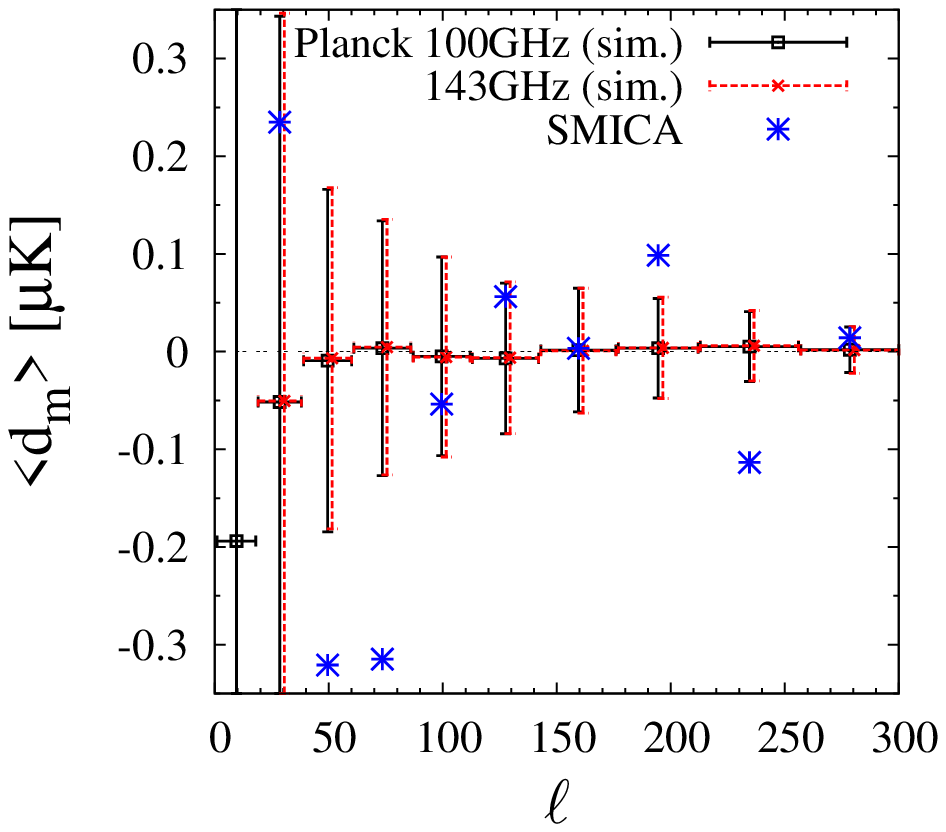}
\end{minipage}
\caption{Histograms (left) and plots (right) of the means $\left<d_m\right>$ calculated using the
 Planck (CMB + Noise) simulations at the $100$ GHz (black) and $143$ GHz (red)
 bands. The means derived from the actual Planck 
 SMICA map are shown as the blue dashed lines (left) and points
 (right). Probability to exceed (PTE) is also shown based on
 the simulations at the $100$ GHz band.}
\label{fig:hist_simulation}
\end{figure}

Figure~\ref{fig:hist_simulation} shows the histograms of the means using
the hundred Planck simulations of CMB with noises at the $100$ GHz and $143$ GHz bands. The
means of the simulations are shown to be consistent with the zero mean
hypothesis, as shown in the right panel of Fig.~\ref{fig:hist_simulation}.  Again, we find the deviations in the multipole ranges
around $\ell\approx 70$ and $\ell \approx 230$ significant. For the
former, we find two samples that show larger deviations from zero
than the actual Planck data, and for the latter, we find no
sample out of the hundred simulations that shows the larger
deviation. The p-values are summarized in table \ref{tb:p-value-simulation}.

\begin{table}[htbp]
\begin{tabular}{ccc}
\hline
\hline
\multicolumn{2}{c}{bin}  &  {$\pm 20^\circ$ cut} \\
$\ell_\mathrm{min}$ & $\ell_\mathrm{max}$ & {Simulation}\\
\hline
19  & 38  &  54\% \\
39  & 60  &  6\% \\
61  & 86  &  2\% \\
87  & 112 &  61\% \\
113 & 142 &  41\% \\
143 & 176 &  96\% \\
177 & 212 &  5\%\\
213 & 256 &  $<$1\% \\
256 & 300 &  58\% \\
\hline
\hline
\end{tabular}
\caption{Probabilities of supporting the null hypothesis ($p$ values)
 that the CMB fluctuations have a zero mean, using the hundred Planck sky simulations. } 
\label{tb:p-value-simulation}
\end{table}

We should caution, however, that the simulations used here do not exactly
correspond to the data we analyze, i.e. the SMICA map. Therefore, these
simulations should be recognized as an estimate of Planck's instrumental
effects in the SMICA map. Furthermore, while we find no sample out of 
the simulations that shows the larger deviation at the $213\lesssim \ell
\lesssim 256$ bin, one should use more simulations to confirm the
significant detection. 

\section{Discussion}
\subsection{Instrument Noise}
\begin{figure}[h]
\begin{minipage}[m]{0.49\textwidth}
\centering
\includegraphics[width=1.0\textwidth]{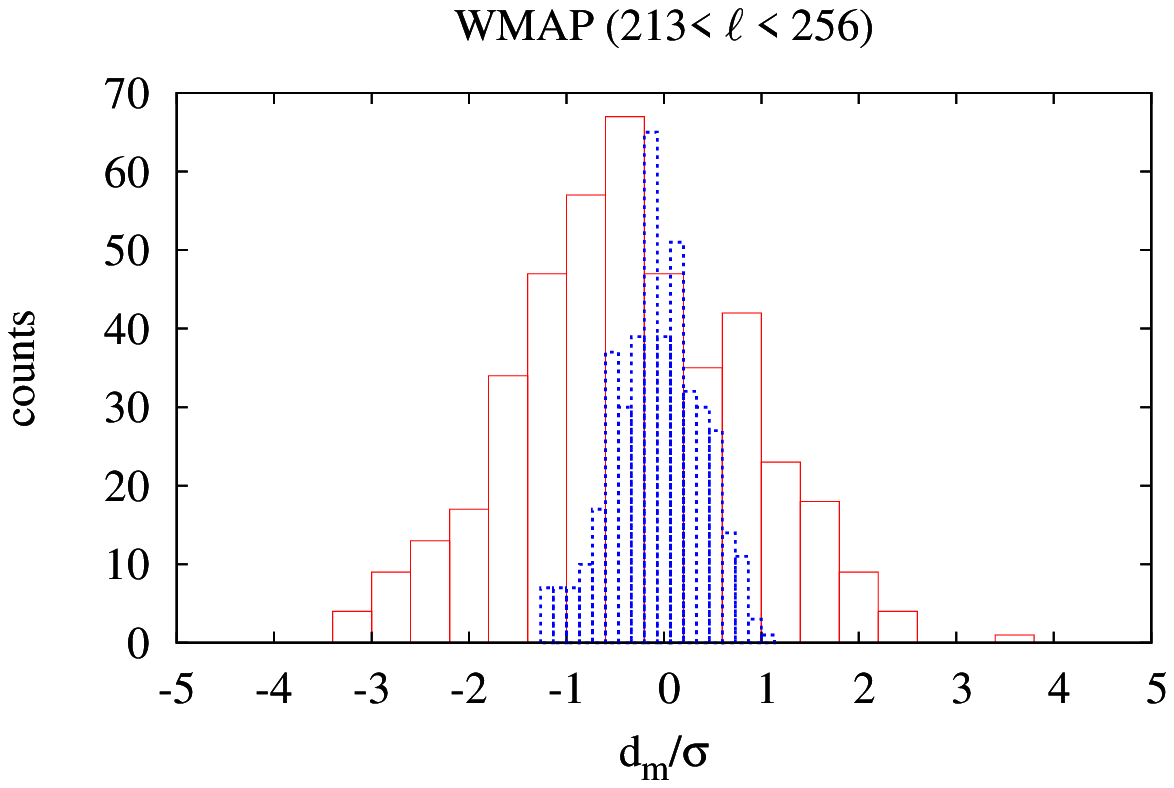}
\end{minipage}
\begin{minipage}[m]{0.49\textwidth}
\centering
\includegraphics[width=1.0\textwidth]{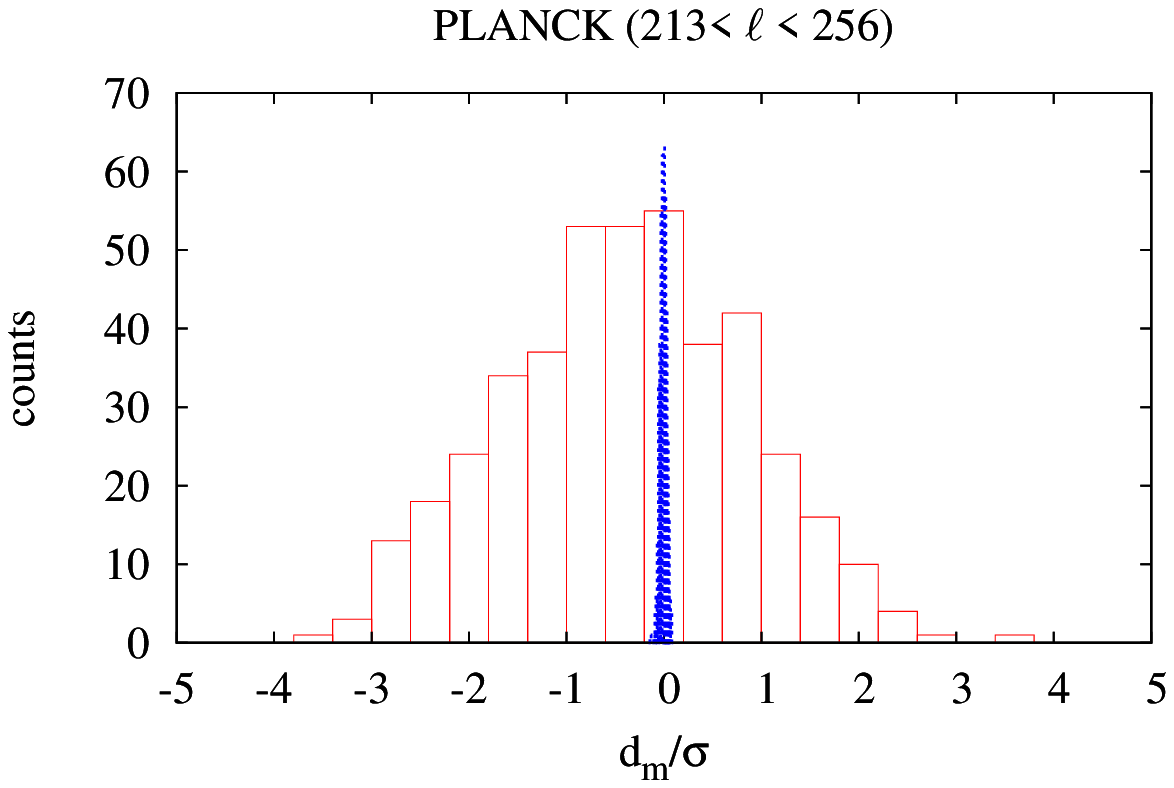}
\end{minipage}
\caption{Histograms of stochastic variable $d_m$ from the signal plus
 noise (red; solid line) and expected noise only (blue; dashed line), normalized by the sample
 variance for the bin of $213\le \ell \le 256$. The top panel is for
 the WMAP data and the bottom is for Planck (SMICA).} 
\label{fig:noise}
\end{figure}

The formulation described in the previous section ignores the effect of
the instrument noise. Although instrument noise that is expected to have
a zero mean would not bias the test of the zero mean hypothesis, it
degrades the statistical power. To demonstrate how the instrument noise
of the WMAP and Planck could affect the test of the zero mean
hypothesis, we construct the variable $d_m$ from the expected noise
values for the sky in the WMAP and Planck SMICA (Spectral Matching
Independent Component Analysis) maps and show the results in
Fig.~\ref{fig:noise} for the multipole range of $\ell\approx
213$-$256$. As expected, the noise in the Planck map are negligibly
small compared with the signal for this angular scale, because of its
high angular resolution. On the other hand, noise can contribute up to
$40$\% for the WMAP case, and this might be a reason for a smaller S/N
from the WMAP than that from Planck at this scale (see
Fig.~\ref{fig:z-score}). Thus we did not explore the test at smaller
scales, $\ell \gtrsim 300$.

\begin{figure}[h]
\centering
\includegraphics[width=0.5\textwidth]{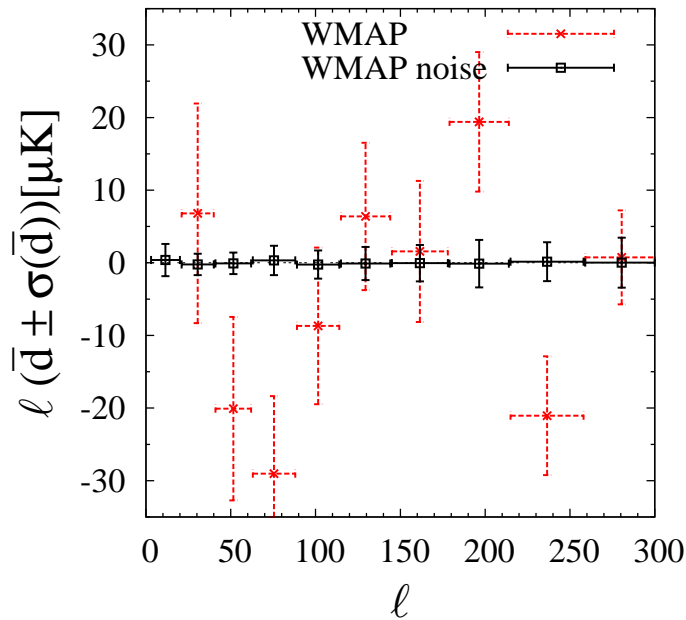}
\caption{Means of stochastic variable $d_m$ from the simulations of the
 WMAP W-band noise (black), together with the actual WMAP data
 (red). The error bars are estimated from the hundred simulations for the
 anisotropic noise, and from the sample for the actual data. We multiply
 y-axis by $\ell$ for visualization purpose.} 
\label{fig:expected_noise}
\end{figure}

While the instrumental noise of the Planck satellite at angular scales
considered here should be significantly lower than the CMB signal as
shown in Fig.~\ref{fig:noise}, that of the WMAP satellite begins to
dominate at highest multipole bins and may bring unwanted statistical
artifacts. Therefore, we make a simple test by simulating the
anisotropic noise based on the hit counts of the WMAP observation.
Specifically, we create hundred noise maps based on the hit count data of
the WMAP W-band observation, apply the same method to the maps, and
obtain the distributions of $d_m$ of the WMAP anisotropic noise.
In Fig.~\ref{fig:expected_noise}, we show the result of our hundred noise
simulation of the WMAP anisotropic noise together with the actual data points.
While, on large angular scales, the WMAP noise is negligibly small compared
with the signal, they become comparable at the $257 \lesssim \ell \lesssim
300$ bin. Thus, as we discussed above, we did not explore the test at
smaller scales, even though the WMAP noise is consistent with zero mean.

While anisotropic noise will have zero mean, the correlated noise in the
time-ordered data could potentially violate the zero mean hypothesis to
some extent, making striping artifacts in the map. However, because the
effect will be different between WMAP and Planck, and therefore we
expect that it is not a major concern given that the results from these
two experiments agree.

\begin{table}[htbp]
\begin{tabular}{ccccccc}
\hline
\hline
\multicolumn{2}{c}{bin}  &  & \multicolumn{4}{c}{$\pm 20^\circ$ cut} \\
$\ell_\mathrm{min}$ & $\ell_\mathrm{max}$ & & SMICA & SEVEM & NILC &
 WMAP \\
\hline
19  & 38  & & 56.1\% & 56.2\% & 56.3\% & 65.3\%   \\
39  & 60  & & 13.1\% & 13.3\% & 13.2\% & 11.2\%  \\
61  & 86  & & 1.38\% & 1.25\% & 1.14\% & 0.659\% \\
87  & 112 & & 58.8\% & 53.1\% & 53.6\% & 42.0\%   \\
113 & 142 & & 46.2\% & 47.8\% & 47.8\% & 52.9\%   \\
143 & 176 & & 95.8\% & 99.6\% & 99.7\% & 87.2\%   \\
177 & 212 & & 5.61\% & 6.11\% & 6.13\% & 4.33\%  \\
213 & 256 & & 0.192\%& 0.213\%& 0.175\% &1.00\%  \\
256 & 300 & & 58.8\% & 64.1\% & 63.1\% & {90.8}\% \\
\hline
\multicolumn{2}{c}{Stouffer's $Z$} & & 1.74 & 1.33 & 1.38 & {1.14}
 \% \\ 
\hline
\hline
\end{tabular}
\caption{Probabilities of supporting the null hypothesis ($p$ values)
 that the CMB fluctuations have a zero mean, for different maps.} 
\label{tb:p-value}
\end{table}

\begin{table}[htbp]
\begin{center}
\begin{tabular}{cccccc}
\hline
\hline
\multicolumn{2}{c}{bin}  &  & $\pm 15^\circ$ & $\pm 20^\circ$ &
 $\pm 25^\circ$ \\
$\ell_\mathrm{min}$ & $\ell_\mathrm{max}$ & & & SMICA &  \\
\hline
19  & 38  & &  59.6\%  &  56.1\%  & 52.7\%  \\
39  & 60  & &  13.1\%  &  13.1\%  & 12.9\%  \\
61  & 86  & &  1.28\%  &  1.38\%  & 1.88\%  \\
87  & 112 & &  63.8\%  &  58.8\%  & 51.6\%  \\
113 & 142 & &  40.1\%  &  46.2\%  & 74.1\%  \\
143 & 176 & &  99.7\%  &  95.8\%  & 92.0\%  \\
177 & 212 & &  5.11\%  &  5.61\%  & 14.4\%  \\
213 & 256 & &  0.511\% &  0.192\% & 0.193\% \\
256 & 300 & &  84.9\%  &  58.8\%  & 28.0\%  \\
\hline
\hline
\end{tabular}
\end{center}
\caption{Probabilities of supporting the null hypothesis ($p$ values)
 that the CMB fluctuations have a zero mean, for different sky cuts. } 
\label{tb:p-value2}
\end{table}

\subsection{Foreground}
\begin{figure}[h]
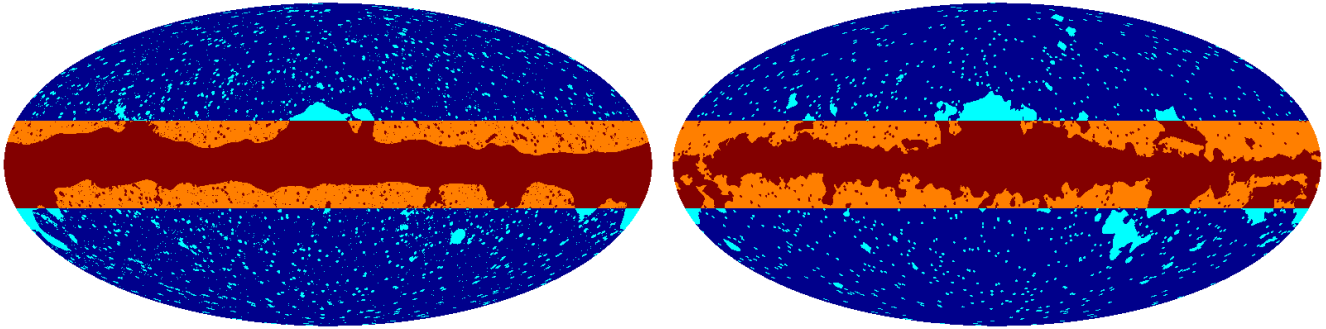

\begin{minipage}[c]{0.49\linewidth}
\includegraphics[width=1.0\textwidth]{mask_comparison.eps3}  
\end{minipage}
\begin{minipage}[c]{0.49\linewidth}
\includegraphics[width=1.0\textwidth]{mask_comparison_wmap.eps3}  
\end{minipage}
\caption{Comparison of the masks used in this paper and cosmological
 analyses by the Planck (left) and WMAP (right) collaborations. The
 masked regions in the Planck and WMAP
 analyses are shown in cyan. Our axisymmetric mask is along the galactic
 plane and shown in orange.} 
\label{fig:3-2}
\end{figure}

\begin{figure}[h]
\begin{minipage}[c]{0.49\linewidth}
\centering
\includegraphics[width=1.0\linewidth]{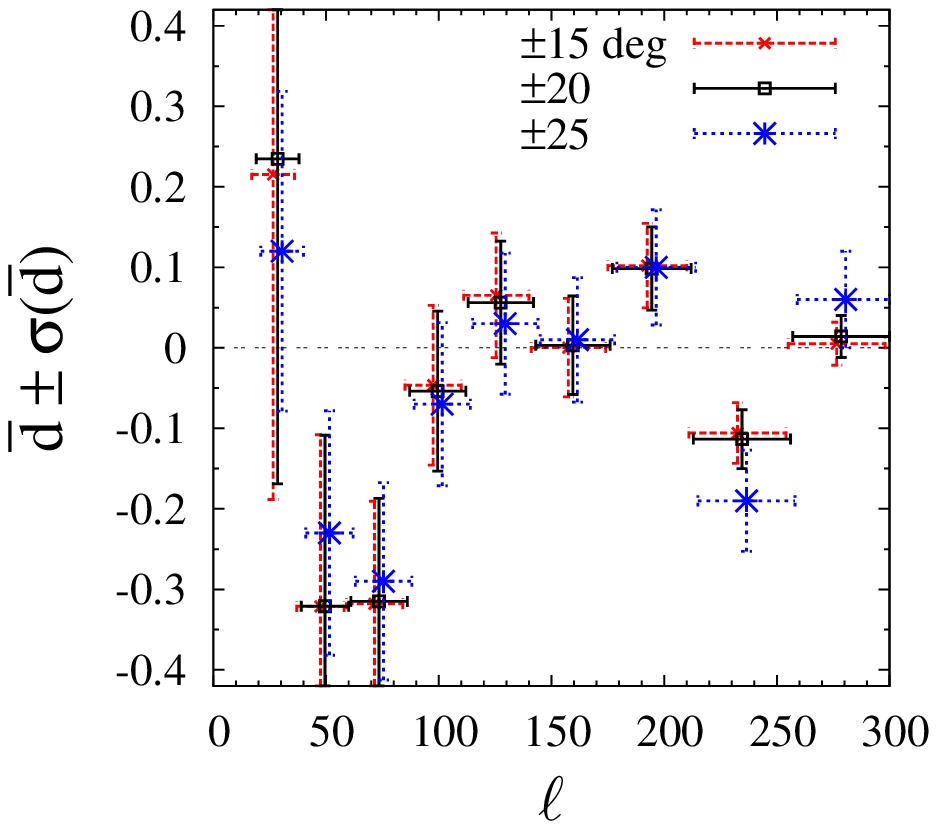}  
\end{minipage}
\begin{minipage}[c]{0.49\linewidth}
\centering
\includegraphics[width=1.0\linewidth]{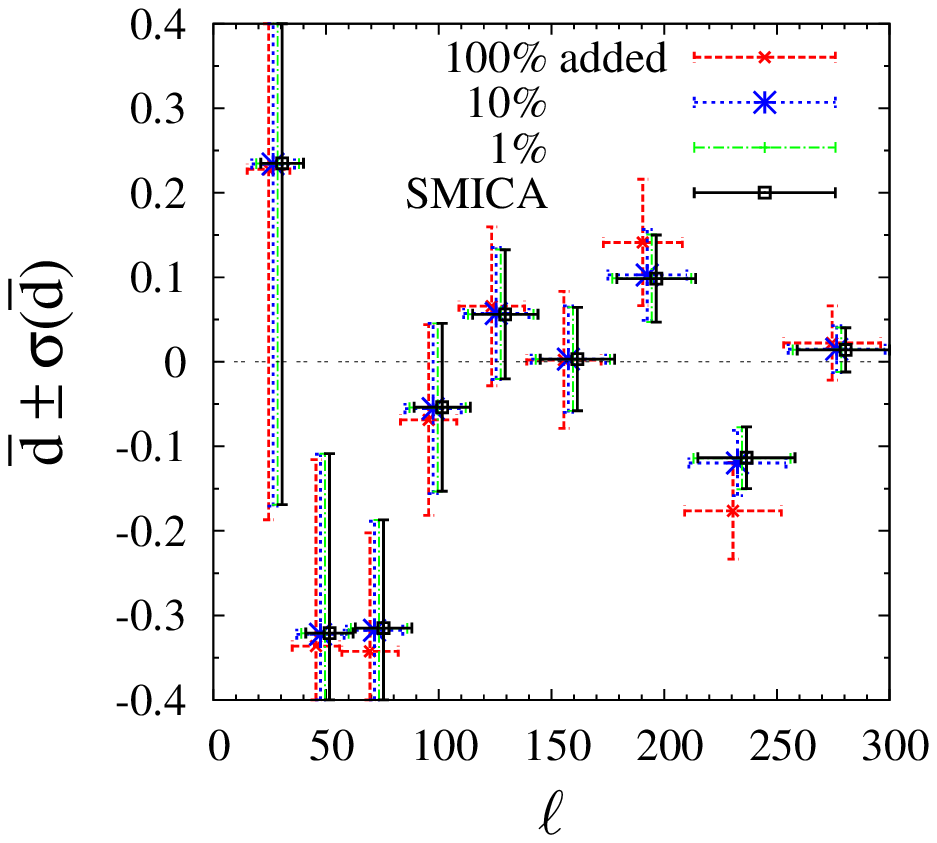}  
\end{minipage}
\caption{Estimated means with error bars for different sky cuts (left)
 and {with different foreground residuals (right)} as
 indicated. In the figure we have not normalized by $\sigma$ as it was done
 in Fig.~\ref{fig:z-score}. We have shifted the $\ell$ values {by $+ 2$
 for $|b|\leq 15^\circ$ and $10$ \% foreground cases, by $-2$ for
 $|b|\leq 25^\circ$ and $1$ \%, and by $+4$ for $100$ \% foreground
 case, for clarity.}}  
\label{fig:4}
\end{figure}

Another issue we have to address is the foreground. One disadvantage of
the simple and clear method described in this paper is that it must
utilize an axisymmetric mask 
to simplify the convolution of the mask as in Eq.~(\ref{eq:3}), 
while the foreground is, of course, not axisymmetric. We compare our
axisymmetric mask with the ones used for the CMB power spectrum
estimates of WMAP and Planck in Fig.~{\ref{fig:3-2}}. Although
directions toward 
galactic disk where the cosmological CMB is heavily contaminated are 
removed by our axisymmetric mask, some parts of the sky at high galactic
latitudes that are removed from the
WMAP and Planck analyses are included in ours.
To estimate how the foreground has contaminated the results in the
previous section, we examine the same analysis but with more extensive
and more aggressive masks that cut the region in the galactic latitude
$|b|\leq 20^\circ \pm 5^\circ$.  The results are shown in
Fig.~{\ref{fig:4}}. Overall, we find mutually consistent results. In
fact, for the eighth bin ($\ell\approx 213 - 256$), the significance
remains the same even for $|b|\leq 25^\circ$ although the standard
deviation $\sigma$ becomes larger because of the smaller analyzed sky
area. This is consistent with what the previous work found with the WMAP
seven-year map \cite{2012PhRvD..85f3001K}.

The same analysis is also done using other types of foreground-reduced
maps from Planck as the SMICA map, namely, SEVEM (internal template
fitting) and NILC (linear internal combination in a needlet space)
maps. These maps are generated through processes completely different
from those of SMICA maps and thus have different weights to both the
frequencies and multipoles.

Table \ref{tb:p-value} and \ref{tb:p-value2} summarize the probabilities
of supporting the null hypothesis for different sky cuts and different
foreground-reduced maps. We find that all the different
foreground-reduced maps from Planck give consistent, almost
indistinguishable results. The WMAP and Planck data are also consistent
with each other, suggesting that instrumental and scanning effects that
may cause apparent violations of statistical homogeneity are
negligible. At the eighth bin ($\ell \approx 213$ - $256$) the signal is
slightly reduced for the WMAP map, but the small signal may be
attributable to the instrument noise as discussed above.  
We also note that while the results in tables \ref{tb:p-value} and
\ref{tb:p-value2} are based on the simple equation
(Eq. (\ref{eq:3.15})), that in table \ref{tb:p-value-simulation} is derived directly from the
histogram of Planck simulations, without assuming Gaussianity. The
results in the tables clearly shows the same tendency.

\begin{figure}[h]
\centering
\includegraphics[width=0.3\textwidth,angle=90]{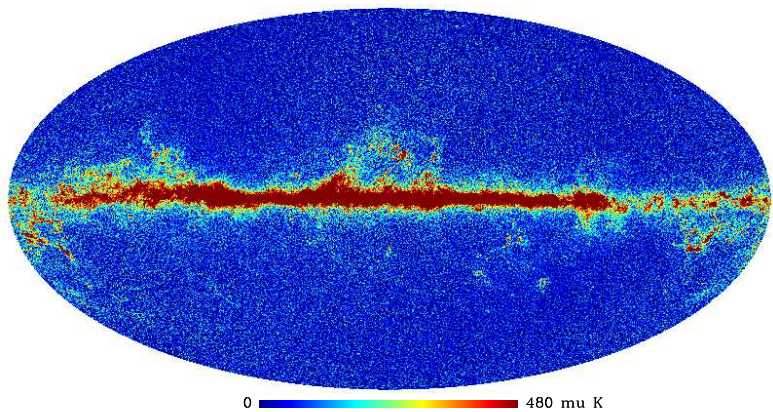}
\caption{The SMICA residual foreground map at the HFI 100GHz band.}
\label{fig:res_smica}
\end{figure}

Even though our result shows that the different foreground-reduced
maps give consistent results, it does not necessarily mean that
foreground contamination is not an issue, because these foreground
subtraction methods are calibrated off the same Planck foreground
model. To see how the foreground residuals could have an effect on the
results, we do a few simple tests with varying amount of residual
foreground contamination using the SMICA residual map at the HFI 100 GHz
band, which is shown in Fig.~\ref{fig:res_smica}. Foreground intensity of 100\%, 10\%, and 1\% of the residual map
are added to the SMICA CMB map and we apply the 
same method to the three maps.  The result is shown in the left panel of
Fig.~\ref{fig:4}. We find that
our results are stable against the residual foreground contamination if the
foreground residuals are less than 10\%.

Another potential issue would be that the zero levels of temperature
(i.e., monopoles) in the Planck experiment are unconstrained. For
example, as discussed in \cite{2013arXiv1303.5066P}, the reported
monopoles with uncertainties are $-300.84 \pm 2.23$, $-22.83 \pm 0.78$
and $-28.09 \pm 0.64$ $\mu$K at the 30, 44, and 70 GHz bands,
respectively. Because it is not trivial how the uncertainty in the
monopole temperature affect the significance of the statistical
inhomogeneity found at high multipole ranges, we do a simple test to
check the effect. Specifically, we apply the same method to the map to
which we add an additional monopole component with $\pm 10$ $\mu$K,
which is much larger than the uncertainty listed above. We found that
the effect can be safely neglected for higher multipole regions
considered in our analysis.

An analysis has been made in Kashino et al. \cite{2012PhRvD..85f3001K},
where we found 
an anomalously large deviation from the zero mean hypothesis at
the multipole range $\ell \approx 221$ - $240$ using Monte-Carlo
simulations with the WMAP maps. The deviation was as large as 99.93\%
confidence level, 
regardless of the different frequency maps and different masks.  
The multipole ranges that show deviations from the zero mean hypothesis
are consistent between Kashino et al. \cite{2012PhRvD..85f3001K} and the
results presented in this 
paper, although the methods used are completely independent from each
other. 


\subsection{Look-Elsewhere Effect}
Finally, let us evaluate the significance as a whole to draw 
a conclusion against the null hypothesis. Because we have tested nine
multipole bins for Planck maps (eight for the WMAP), we should take into
account the effect whether apparent anomalies are found just because of 
statistical outliers. The effect is often called the look-elsewhere effect.
In order to take this effect into account, we combine $p$-values by
calculating the Stouffer's weighted $Z$ (Liptak-Stouffer method), which
is defined as \cite{2005Whitlock} 
\begin{equation}
 Z\equiv
\frac{\sum_1^n w_i Z_i}
{\sqrt{\sum_1^n w_i^2}}~.
\end{equation}
Here $Z_i$ is the so-called $Z$-score defined by $Z_i=\Phi^{-1}(1-p_i)$,
where $\Phi$ is the standard normal cumulative distribution function 
and $w_i$ is the number of degrees of freedom for the $i$-th bin.  The
combined variable $Z$, which follows the standard normal distribution if
the common hypothesis is true, reflects the fact that we have done
multiple tests for a common hypothesis. From the p-values listed in
Table \ref{tb:p-value}, we find the values of Stouffer's weighted Z as
\begin{equation}
 Z = 1.74~\mbox{(SMICA)},~1.14~\mbox{(WMAP)}~,
\end{equation}
which means, the percentages to reject the zero mean hypothesis are 
$91.8$ \% and $74.5$ \%, respectively.
Therefore we may conclude that the zero mean hypothesis is consistent
with observational data as a whole.

There is the possibility to perform additional tests to confirm whether
or not the anomalous
deviations from the zero mean hypothesis found here are just 
statistical fluctuations due to a particular realization of the
Universe.  A straightforward test is to make use of full-sky CMB 
polarization data that will soon be released from the Planck collaboration.
Although the polarization anisotropies are made from  common
curvature fluctuations, their transfer functions do not completely
coincide with those of temperature anisotropies, and thus they will lend 
additional statistical power. Another test is to look into large-scale
structure data, which offers an independent probe for primordial
fluctuations
\cite{2011CQGra..28p4003S,2013ApJ...762L...9H,2013MNRAS.430.3376P}.  
The comoving scale that corresponds to the multipole range
of $\ell \approx 213$ - $256$ is approximately $k\approx 0.015$ - $0.018$
Mpc$^{-1}$, which is at the edge of the current galaxy survey by BOSS
\cite{2013arXiv1312.4611B} 
and will be within reach in future galaxy surveys, such as
Euclid \cite{2011arXiv1110.3193L}, LSST \cite{2009arXiv0912.0201L}, 
SKA \footnote{https://www.skatelescope.org} and others.
Interesting ideas have been discussed in
Refs.~\cite{2011JCAP...09..035H,1995PhRvD..52.1821G,2008PhRvL.100s1302C,2012PhRvL.109e1303C,2013ApJ...762L...9H},
which include arguments that 
cosmic star formation histories and the kinetic Sunyaev-Zel'dovich effect can be
used to probe inside our past light cone and thus they become powerful
tools to probe into the cosmic homogeneity.

Before concluding, we would like to comment on the connection
with studies on non-Gaussianity in the CMB. In analyses of higher-order
statistics, such as the 
bispectrum, the zero mean condition has been implicitly  
assumed and one estimates a correlation of the form $\left<a_{\ell m}
a_{\ell^\prime m^\prime} a_{\ell'' m''}\right>$. Consider a case where
the mean of $a_{\ell 
m}$ was not zero but the bispectrum was zero around the mean; it is expected
that the three point correlation of the above form would have an amplitude on the order of 
\begin{equation}
\left<a_{\ell m}
a_{\ell^\prime m^\prime}
a_{\ell'' m''}
 \right>
\sim
C_\ell \left<{a}_{\ell m}\right>.
\end{equation}
Therefore, constraints on non-Gaussianity using the bispectrum in the
literature could be used to put constraints on the mean of the spherical
harmonic coefficients when this is the case.

\section{conclusion}
We have tested one working cosmological hypothesis, which states that
cosmological perturbations have a zero ensemble mean, using the latest CMB 
temperature anisotropy maps from the WMAP and Planck satellites. We find
evidence against the zero mean hypothesis in two particular ranges of
multipoles, with significance levels of $2.5 \sigma$ at $\ell \approx
61$ - $86$ and $3.1 \sigma$ at $\ell \approx 213$ - $256$. However, in the
present analysis, we conclude that the zero mean hypothesis is consistent
with the current observational data on the basis of the Stouffer's weighted-Z
statistics, which takes into account multiple testing. The zero mean
hypothesis can be further tested by future CMB polarization data that
will be available soon from
Planck satellite.

\acknowledgments 
The author would like to thank T. T. Takeuchi, D. Kashino, M. Sasaki, T.
Tanaka, M. Butcher, and A. Taruya for useful discussions especially at
the Mini-workshop on Gravitation and Cosmology for APC-YITP
collaboration.  Thanks also go to the anonymous referees for giving us
helpful and valuable comments.
This work has been supported in part by Grant-in-Aid for
Scientific Research No. 24340048 from the Ministry of Education, Sports,
Science and Technology (MEXT) of Japan.


\bibliography{paper}

\end{document}